\documentstyle[12pt,epsfig]{article}
\begin{document}
\title{Aharonov--Bohm Effect and Coordinate Transformations.}
\author{ A. Camacho
\thanks{email: acamacho@aip.de} \\
Astrophysikalisches Institut Potsdam. \\
An der Sternwarte 16, D--14482 Potsdam, Germany.}

\date{}
\maketitle

\begin{abstract}
Resorting to a {\it Gedankenexperiment} which is very similar to the famous Aharonov--Bohm proposal 
it will be shown that, in the case of a Minkowskian spacetime, we may use a nonrelativistic quantum 
particle and a noninertial coor\-di\-na\-te system and obtain geometric information of 
regions that are, to this particle, for\-bi\-dden. 
This shows that the outcome of a nonrelativistic quantum process is determined not only by the features of 
geo\-me\-try at those points at which the process takes place, but also by geo\-me\-tric parameters of regions in which the quantum system can not enter. 
From this fact we could claim that geometry at the quantum level plays a non--local role. 
Indeed, the measurement outputs of some nonrelativistic quantum experiments 
are determined not only by the geometry of the region in which the experiment takes place, but 
also by the geometric properties of spacetime volumes which are, in some way, forbidden in the experiment. 

\end{abstract} 
\newpage
\section{Introduction.}
\bigskip

Quantum Theory (QT) has become one of the most successful human achievements, and 
almost all of physics now relies upon QT. Nevertheless, there are some old 
conceptual puzzles that still beset this theory. 
For instance, the so called quantum measurement problem (the problem of the quantum limit) [1], 
the possible incompleteness of the general--relativistic description of gravity in the context of QT [2], 
or the possible discrepancy, in a curved manifold, between Feynman and Schr\"odinger 
formalisms [3].
    
At the classical level gravity can be understood as a purely geometric effect, the 
motion of a free classical particle moving in a curved manifold is given by the Weak Equivalence Principle (WEP), i.e., 
the particle moves along geodesics. The inclusion of additional interactions is done 
resorting to SEP, the famous ``semicolon goes to coma rule'' [4]. 
This principle tells us that locally the laws of physics are the special--relativistic laws.

Classically the role of geometry is local, i.e., the dynamics of a free cla\-ssi\-cal particle located at 
a certain point $P$ of any Riemannian manifold is, according to General Relativity (GR), 
determined by the geometric properties of this manifold at $P$ (the motion equations 
can be written in terms of the connection coefficients, which at $P$ depend only on the values 
of the components of the metric and their derivatives evaluated at $P$), geometry at any other point 
plays no role in the determination of the motion when the particle is at $P$. 
If we consider the geodesic deviation between two particles, then we would obtain 
information of Riemann tensor, but once again only of the region where the motion of these 
classical particles takes place.

Nevertheless, at the quantum level the situation could be not so satisfactory. 
Indeed, the  experiment of Colella, Overhauser and Werner [5] 
tells us that at the quantum level gravity is not anymore a purely geometric effect [6], 
the mass of the employed particles appears explicitly in the interference term. This 
fact emerges once again if we measure continuously the position of the particles [7] 
(even if the particles follow the same trajectory), and could lead to the emergence, 
in some cases, of something like a gravitational quantum Zeno effect [8].

In order to understand better the possible appearance of nonlocal effects in QT let us at this 
point address an already known similarity between electrodynamics and gravitation. 

At the classical level the motion of a charged particle is solely determined by the 
force law of Lorentz and Newton's second law. We also know that the electric and magnetics 
fields are invariant under the so called gauge transformations [9].

Nevertheless, this is not the situation at the quantum level. The role 
that the concept of potential plays in physics has been deeply modified 
by Aharonov--Bohm effect (AB) [10]. Indeed, even though Lorentz force vanishes at those points at which the wave function 
has nonvanishing values, the dynamical behavior is sensitive to the existence of a magnetic field 
inside a region where the charged particle can never enter, the vector potential 
${\bf A}$ has in this effect a measurable consequence, 
detectable in the interference pattern of a charged particle. AB shows us also 
that in QT there are nonlocal effects, i.e., the features of the vector potential at points where 
the wave function vanishes affects the dynamics of the particle, in other words, we could say that in QT ${\bf A}$, sometimes, has a nonlocal role. This does not happen in 
the classical case, where forces have a local character. This effect has already been confirmed experimentally [11].  

In classical physics ${\bf A}$  is a gauge field, it has no physical meaning (at least before a gauge is imposed), it is 
the field strength which is physically relevant. Concerning geometry something similar happens in relation with the 
components of the metric tensor, $g_{\mu\nu}$. They are deprived of physical meaning (of course, 
the metric tensor has physically relevant meaning, but not its components), and sometimes play also 
the role of a potential, i.e., gravitational potential. Therefore we may wonder if we could find a 
construction in which (in analogy with the electrodynamical case, where a nongauge invariant 
field renders nonlocal effects in QT) 
these noninvariant (under coordinate transformations) parameters, $g_{\mu\nu}$, could allow us to find nonlocal effects in 
QT. It is in this sense that here we will speak of a coordinate transformations--induced 
Aharonov--Bohm effect, the appearance of a nonlocal behavior in nonrelativistic QT by means of coordinate transformations.
   
In this work we will consider a Minkowskian spacetime and two coordinate systems in it. 
One of them is an inertial system, while the second one is accelerated. We will prove 
resorting to a {\it Gedankenexperiment}, which is very similar to the AB case, 
that in the accelerated system nonlocal effects could appear in the context of nonrelativistic QT, and that this 
is a geometry--induced feature.
\bigskip

\section{Transformations and Aharonov--Bohm Effect.}
\bigskip

Consider a Minkowskian spacetime, and let us denote the coordinates of an inertial 
coordinate system by $x^{\tilde{\alpha}}$. The matrix elements 
of the coordinate transformation leading to a second coordinate system (which in general 
is noninertial, and whose coordinates will be denoted by $x^{\beta}$) are given by $\Lambda^{\beta}_{\tilde{\alpha}}$. In other words, 
we have  $x^{\beta} = \Lambda^{\beta}_{\tilde{\alpha}}x^{\tilde{\alpha}}$. Notice that 
no conditions have been imposed upon $\Lambda^{\beta}_{\tilde{\alpha}}$.

Let us now proceed to analyze the movement of a quantum particle, and denote its corresponding Lagrangian by $L$. 
We will restrict ourselves to the case of low velocities, i.e., velocities much 
smaller than the speed of light. 

The motion of a free classical particle in a Riemannian manifold is given by the co\-rres\-pond\-ing geodesics [4]. 
In the case of a Minkowskian spacetime, an inertial coor\-di\-na\-te system obtains the 
motion equations calculating the extremal curves of the following expression

\begin{equation}
S= \int (-\eta_{\tilde{\mu}\tilde{\nu}}{dx^{\tilde{\mu}}\over d\tau}{dx^{\tilde{\nu}}\over d\tau})^{1/2}d\tau.
\end{equation}
\bigskip

But in the noninertial system one has to consider not expression (1) but 

\begin{equation}
S= \int (-g_{\mu\nu}{dx^{\mu}\over d\tau}{dx^{\nu}\over d\tau})^{1/2}d\tau.
\end{equation}
\bigskip

From (2) we deduce the motion equations in the non--inertial system

\begin{equation}
{d^2x^{\beta}\over d\tau^2} + \Gamma^{\beta}_{\mu\nu}{dx^{\mu}\over d\tau}{dx^{\nu}\over d\tau} = 0,
\end{equation}
\bigskip

\noindent here $\Gamma^{\beta}_{\mu\nu}$ are the so called connection coefficients, and $\tau$ represents
proper time.

In other words, we may interpret expression (2) as the action of a classical particle in the noninertial system, 
and $L = (-g_{\mu\nu}{dx^{\mu}\over d\tau}{dx^{\nu}\over d\tau})^{1/2}$ as its Lagrangian.

We now introduce in this Minkowskian spacetime a very particular coordinate system. 

Consider a cylindrical volume $H$ (this volume is infinitely long, and its cross section, denoted by $E$, has radius $\rho_a$, see figure). 
Let us now introduce a vector field, denoted by ${\bf A}$, such that it satisfies the following conditions: 
(i) inside $H$ (if $0 \leq \rho \leq \rho_a$) we have, in cylindrical coordinates, $A_{\rho} = 0$, 
$A_{z} = 0$, and $A_{\phi} = F\rho/2 $; (ii) outside $H$ (if $\rho_a \leq \rho$), 
$A_{\rho} = 0$, $A_{z} = 0$, and finally $A_{\phi} = F\rho_a/2$. 
Here $\rho_a > 0$ is a fixed number and $F$ is a nonvanishing real number. Clearly, ${\bf A}$ is everywhere continuous. 
From this definition we may evaluate its rotational; $\nabla\times{\bf A} = {\bf 0}$, if $\rho_a \leq \rho$, and 
$\nabla\times{\bf A} = F\hat{{\bf z}} = {\bf F}$, if $0 \leq \rho \leq \rho_a$, being $\hat{{\bf z}}$ the unit vector along the axis 
of symmetry.

We may now define, using vector field ${\bf A}$, the components of the metric of our accelerated coordinate system, 
namely $g_{0\phi,\rho} = A_{\phi}$, $g_{0z,\rho} = A_{z}$, $g_{0\rho,\rho} = A_{\rho}$, 
$g_{00,\nu} = 0$, and $ g_{0l,j} = 0$, if $j \not = \rho$ (here $l$ and $j$ represent space coordinates, 
while $\nu$ denotes spacetime ones).

The mathematical consistency of this noninertial metric is determined by the existence 
of a coordinate transformation that could render the aforesaid conditions. In order to 
see that we may have this kind of coordinate system, we must note that in this situation 
we must determine 16 functions, namely 
$\Lambda^{\tilde{\beta}}_{\mu}$ (because we have differential 
equations in terms of the components of the metric of the accelerated system, and we also know that 
$g_{\mu\nu} = \Lambda^{\tilde{\beta}}_{\mu}\Lambda^{\tilde{\alpha}}_{\nu}\eta_{\tilde{\beta}\tilde{\alpha}}$). 
But we have only 13 equations, and in consequence the system, in principle, is solvable.

We now construct a {\it Gedankenexperiment} that could be considered, in some way, an extrapolation 
of the famous Aharonov--Bohm construction [10].

Take two points $P$ (source point) and $Q$ (detection point) in this manifold such that the above mentioned cylindrical volume lies between them (see figure).

A particle will move from $P$ to $Q$. It first passes through a conventional 
two--slit device (here we consider each slit as a finite ``hole'') [12], and afterwards enters a 
region in which a forbidden volume $D$ for this particle exists (this volume $D$ is infinitely 
long, and contains in its interior the, also infinitely long, cylinder $H$ in which 
$\nabla\times {\bf A} \not ={\bf 0}$). 
Then it is detected at point $Q$. 
In other words, after passing the two--slit device it remains always on one ``side'' of space, either ``right'' or ``left'', 
volume $D$ acts as a barrier for the particle.

Under these conditions the proper time of any curve $C$ joining $P$ and $Q$ is given by 

\begin{equation}
S= \int_C (-g_{\mu\nu}{dx^{\mu}\over d\tau}{dx^{\nu}\over d\tau})^{1/2}d\tau.
\end{equation}
\bigskip

Let us now suppose that the source and detection points are moved along the $\rho$ coordinate 
a distance $-\epsilon$, i.e., any curve $x^{\alpha} = C^{\alpha}(\tau)$ joining $P$ and $Q$ will become 
now $x^{\alpha} = C^{\alpha}(\tau)$ for $\alpha \not =\rho$, and $\rho(\tau) = C^{\rho}(\tau) - \epsilon$, with the condition $|\epsilon| << 1$. 

In this new situation expression (4) becomes

\begin{equation}
S= \int_C (-g_{\mu\nu}{dx^{\mu}\over d\tau}{dx^{\nu}\over d\tau} + 
\epsilon{\partial g_{\mu\nu}\over \partial \rho}{dx^{\mu}\over d\tau}{dx^{\nu}\over d\tau})^{1/2}d\tau.
\end{equation}
\bigskip

We now consider a quantum particle moving (here only the case of low velocities is analyzed) from the new source point to the new detection point.
The description of the movement of this particle can be done using Feynman's 
path integral formulation for a nonrelativistic particle [13], thus its propagator $U$ is given by

\begin{equation}
U({\bf x}_2, \tau ''; {\bf x}_1, \tau ') = \int d[{\bf x}(\tau)]exp\Bigl({i\over\hbar}\int_{\tau '}^{\tau ''}
(-g_{\mu\nu}{dx^{\mu}\over d\tau}{dx^{\nu}\over d\tau} + 
\epsilon{\partial g_{\mu\nu}\over \partial \rho}{dx^{\mu}\over d\tau}{dx^{\nu}\over d\tau})^{1/2}d\tau\Bigr),
\end{equation}

being ${\bf x}_2$ and ${\bf x}_1$ the space coordinates of the new detection point and 
of the new source point, respectively.

But low velocities (${dt\over d\tau} \sim 1$, here $c =1$) and $g_{00, \rho} = 0$ imply that the propagator is approximately

\begin{equation}
U({\bf x}_2, \tau ''; {\bf x}_1, \tau ') = \int d[{\bf x}(t)]exp\Bigl({i\over\hbar}\int_{\tau '}^{\tau ''}
(L + \epsilon A_l{dx^{l}\over dt})dt\Bigr),
\end{equation}

\noindent being $L= (-g_{\mu\nu}{dx^{\mu}\over dt}{dx^{\nu}\over dt})^{1/2}$. 
\bigskip

\section{Interference Terms.}
\bigskip

Let us now calculate the probability of detecting our particle. 
Clearly, the propagator at this point is the sum of two terms, the propagator ``right'' and the propagator ``left''.

{\setlength\arraycolsep{2pt}\begin{eqnarray}
U({\bf x}_2, \tau ''; {\bf x}_1, \tau ') = \int_{(right)}d[{\bf x}(t)]exp\Bigl({i\over\hbar}\int_{\tau '}^{\tau ''}[L + \epsilon A_l{dx^{l}\over dt}]dt\Bigr) \nonumber\\
+ \int_{(left)}d[{\bf x}(t)]
exp\Bigl({i\over\hbar}\int_{\tau '}^{\tau ''}[L + \epsilon A_l{dx^{l}\over dt}]dt\Bigr).
\end{eqnarray}} 
\bigskip

 As was mentioned before, in this {\it Gedankenexperiment} the particles can not go from the 
left--hand side to the right--hand side (or from the right--hand side to the left--hand side). 
These two conditions imply that in our case orbits are confined to a topologically restricted 
part of space.

If we carry out the so called ``skeletonization'' , then we may see that the contribution 
to the respective integrals of each trajectory can be written as follows

\begin{equation}
\prod_{n =1}^{N-1}exp\Bigl({i\over\hbar}S[n, n+1] + {i\over\hbar}T_{top}\Bigr)exp\Bigl(\epsilon{i\over\hbar}\int_{C}{\bf A}\cdot d{\bf s}\Bigr),
\end{equation}
\bigskip

\noindent being $C$ the trajectory under consideration joining new source point and new detection point, $s^l = {dx^{l}\over dt}dt$, and $S$ the action associated to $L$. 
The new term $T_{top}$ is a pure boundary term, which keeps track of the imposed topological restrictions [14]. 
Either ``right'' or ``left'' the rotational of ${\bf A}$ vanishes, therefore the 
line integral on the last term of expression (9) depends only on the initial and final points, and not on $C$ 
(it is readily seen that $C$ is not a closed curve, and that it lies outside cylinder $H$). 
In other words, if we consider two trajectories ``right'' (``left'') the contribution 
of our vector field ${\bf A}$ to each one of them is the same, i.e., the exponential of the line integral of ${\bf A}$ is a common factor.

Hence the propagator becomes now

{\setlength\arraycolsep{2pt}\begin{eqnarray}
U({\bf x}_2, \tau ''; {\bf x}_1, \tau ') = exp\Bigl(\epsilon{i\over\hbar}\int_{C1}{\bf A}\cdot d{\bf s}\Bigr)\int_{(right)}d[{\bf x}(t)]
exp\Bigl({i\over\hbar}\int_{\tau '}^{\tau ''}\tilde{L}dt\Bigr) \nonumber\\
+~exp\Bigl(\epsilon{i\over\hbar}\int_{C2}{\bf A}\cdot d{\bf s}\Bigr)\int_{(left)}d[{\bf x}(t)]
exp\Bigl({i\over\hbar}\int_{\tau '}^{\tau ''}\tilde{L}dt\Bigr).
\end{eqnarray}} 
\bigskip

Here $C1$ and $C2$ are any trajectory (joining points $P$ and $Q$) ``right'' and ``left'', respectively, and $\tilde{L}$ represents 
the Lagrangian function $L$ plus the topological term $T_{top}$.

From expression (10) we may evaluate the in\-ter\-fe\-rence term.

\begin{equation}
I = 2\alpha cos\Bigl(\epsilon{1\over\hbar}\oint_{\tilde{C}}{\bf A}\cdot d{\bf s}\Bigr) - 
2\beta sin\Bigl(\epsilon{1\over\hbar}\oint_{\tilde{C}}{\bf A}\cdot d{\bf s}\Bigr).
\end{equation} 
\bigskip

In expression (11) the closed curve $\tilde{C}$ is defined with $C1$ and $C2$. Firstly, we move from 
the new source point to the new detection point along $C1$, and then backwards along $C2$. We have also introduced the following definitions 

\begin{equation}\alpha = Re\{\int_{(right)}d[{\bf x}(t)]
exp\Bigl({i\over\hbar}\int_{\tau '}^{\tau ''}\tilde{L}dt\Bigr)\int_{(left)}d[{\bf x}(t)]
exp\Bigl(-{i\over\hbar}\int_{\tau '}^{\tau ''}\tilde{L}dt\Bigr)\},  
\end{equation} 
\bigskip

and 

\begin{equation}\beta = Im\{\int_{(right)}d[{\bf x}(t)]
exp\Bigl({i\over\hbar}\int_{\tau '}^{\tau ''}\tilde{L}dt\Bigr)\int_{(left)}d[{\bf x}(t)]
exp\Bigl(-{i\over\hbar}\int_{\tau '}^{\tau ''}\tilde{L}dt\Bigr)\}.  
\end{equation} 
\bigskip

Using Stokes' theorem we may rewrite this interference term as

\begin{equation}
I = 2\alpha cos\Bigl(\epsilon{1\over\hbar}\int_\Omega \nabla\times{\bf A}\cdot d{\bf \Omega}\Bigr) - 
2\beta sin\Bigl(\epsilon{1\over\hbar}\int_\Omega \nabla\times{\bf A}\cdot d{\bf \Omega} \Bigr),
\end{equation} 
\bigskip

being $\Omega$ an area bounded by $\tilde{C}$. But we have defined our vector field ${\bf A}$ 
such that its rotational vanishes everywhere on $\Omega$ but in a small area located inside the forbidden volume $D$, 
i.e., in the cross section of cylinder $H$, which was denoted by $E$ and has radius $\rho_a$. 
Hence the nonvanishing part of the line integrals allows us to rewrite (14) as follows

\begin{equation}
I = 2\alpha cos\Bigl(\epsilon{1\over\hbar}\int_E \nabla\times{\bf A}\cdot d{\bf E}\Bigr) - 
2\beta sin\Bigl(\epsilon{1\over\hbar}\int_E \nabla\times{\bf A}\cdot d{\bf E} \Bigr).
\end{equation} 
\bigskip

From our previous definitions we obtain the final form of this interference term 

\begin{equation}
I = 2\alpha cos\Bigl(\epsilon{F\over\hbar} \pi\rho_a^2\Bigr) - 
2\beta sin\Bigl(\epsilon{F\over\hbar} \pi\rho_a^2\Bigr).
\end{equation} 

\bigskip
\bigskip

\section{Discussion.}
\bigskip

We have proved, using a {\it Gedankenexperiment} which is very similar to the famous Aharonov--Bohm proposal, 
that we may find noninertial coordinate systems (in a Minkowskian spacetime), in which 
a nonrelativistic quantum process is determined not only by the features of geometry at those points at which 
the process takes place, but also by geometric parameters of regions in which the quantum system can not enter. 

This is a purely quantum mechanical effect. Indeed, if we had used a classical particle, 
where the motion at any point $P$ of its trajectory is solely determined by the geometry at $P$, then 
no information of any forbidden region could be extracted.  

The here introduced {\it Gedankenexperiment} could hardly be considered as a local experiment, nevertheless, from our results 
we could claim that geometry--induced nonlocal effects could emerge in QT. Indeed, the measurement outputs of some nonrelativistic quantum experiments 
are determined not only by the geometry of the region in which the experiment takes place, but 
also by the geometry of regions forbidden, in some way, to the experiment. 

The gravitational field can be geometrized (at least at the classical level), and at this point we may wonder if at quantum 
level gravity could render some nonlocal e\-ffects in nonrelativistic QT (of course, the 
present work does not consider gravity, but it has proved that in a Minkowskian spacetime 
there could be a geometry--induced nonlocality, and therefore the extension to curved spacetimes is an interesting question). 
This nonlocal behavior has already been pointed out [15], and more investigation around this topic could lead to a more profound comprehension 
of the way in which the gravitational field could modify some fundamental expressions of QT, for example the commutation relations [16, 17]. 

The possible incompleteness of the general relativistic description of gravity, 
at quantum level, has already been claimed [2], and implies the 
violation not only of Einstein equivalence principle but also of the local position 
invariance principle [18] (the independence of the results of local experiments from 
the location of the local laboratory in spacetime, i.e., the independence of the 
equivalence principle from position in time and space). In other words, 
Ahluwalia's work implies that the results of some local quantum experiments do depend on nonlocal characteristics. 
In the present work we have found a behavior that, at least qualitatively, is the same, i.e., 
the dynamics of some quantum processes is determined by nonlocal features. Hence, further 
investigation in this Aharonov--Bohm effect could help to understand better the controversy around the validity, at quantum level, 
of the equivalence principle

In the present work, the physical acceptability of the constructed noninertial system 
has not been investigated. In spite of this last fact, our work has shown that, in principle, 
there are noninertial systems, in which nonrelativistic QT shows very interesting nonlocal features. The possibility 
of having also these kind of effects in the context of more realistic schemes (feasible 
noninertail observers) has to be investigated, but at least we have shown that in a very wide 
range of noninertial coordinate systems these effects do exist.

\bigskip

\Large{\bf Acknowledgments.}\normalsize
\bigskip

The author would like to thank A. Camacho--Galv\'an and A. A. Cuevas--Sosa for their 
help, and D.-E. Liebscher for the fruitful discussions on the subject. 
The hospitality of the Astrophysikalisches Institut Potsdam is also kindly acknowledged. 
This work was supported by CONACYT Posdoctoral Grant No. 983023.

\setlength{\unitlength}{0.00083300in}%
\begingroup\makeatletter\ifx\SetFigFont\undefined
% extract first six characters in \fmtname
\def\x#1#2#3#4#5#6#7\relax{\def\x{#1#2#3#4#5#6}}%
\expandafter\x\fmtname xxxxxx\relax \def\y{splain}%
\ifx\x\y   % LaTeX or SliTeX?
\gdef\SetFigFont#1#2#3{%
  \ifnum #1<17\tiny\else \ifnum #1<20\small\else
  \ifnum #1<24\normalsize\else \ifnum #1<29\large\else
  \ifnum #1<34\Large\else \ifnum #1<41\LARGE\else
     \huge\fi\fi\fi\fi\fi\fi
  \csname #3\endcsname}%
\else
\gdef\SetFigFont#1#2#3{\begingroup
  \count@#1\relax \ifnum 25<\count@\count@25\fi
  \def\x{\endgroup\@setsize\SetFigFont{#2pt}}%
  \expandafter\x
    \csname \romannumeral\the\count@ pt\expandafter\endcsname
    \csname @\romannumeral\the\count@ pt\endcsname
  \csname #3\endcsname}%
\fi
\fi\endgroup
\begin{picture}(7449,8460)(2389,-7888)
\thicklines
\put(2401,-2161){\line( 1, 0){1800}}
\put(4426,-2161){\makebox(6.6667,10.0000){\SetFigFont{10}{12}{rm}.}}
\put(4501,-2161){\line( 1, 0){3300}}
\put(8026,-2161){\line( 1, 0){1800}}
\put(4801,-2161){\line( 0,-1){5025}}
\put(4801,-7186){\line( 1, 0){2925}}
\put(7726,-7186){\line( 0, 1){  0}}
\put(7726,-7186){\line( 0, 1){5025}}
\put(6001,-5611){\circle{1060}}
\put(6001,-61){\makebox(0,0)[lb]{\smash{\SetFigFont{12}{14.4}{rm}P}}}
\put(6001,-5611){\makebox(0,0)[lb]{\smash{\SetFigFont{12}{14.4}{rm}E}}}
\put(6001,-7561){\makebox(0,0)[lb]{\smash{\SetFigFont{12}{14.4}{rm}Q}}}
\put(5401,-7861){\makebox(0,0)[lb]{\smash{\SetFigFont{12}{14.4}{rm}Detection Point.}}}
\put(5401,-3136){\makebox(0,0)[lb]{\smash{\SetFigFont{12}{14.4}{rm}Forbidden Volume D}}}
\put(8401,-5161){\makebox(0,0)[lb]{\smash{\SetFigFont{12}{14.4}{rm}``Right''}}}
\put(3826,-5161){\makebox(0,0)[lb]{\smash{\SetFigFont{12}{14.4}{rm}``Left''}}}
\put(5401,464){\makebox(0,0)[lb]{\smash{\SetFigFont{12}{14.4}{rm}Source Point.}}}
\end{picture}

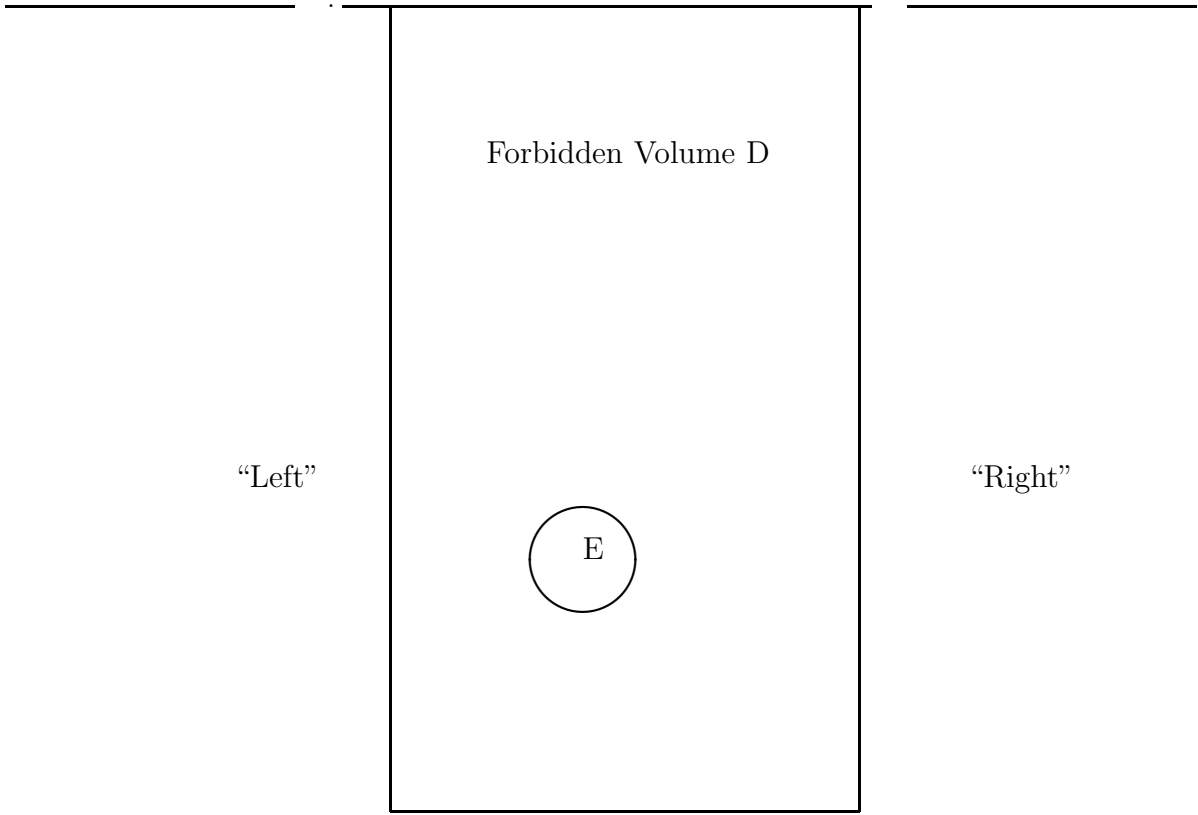
\end{document}